\documentclass[aps,pra,amsfonts,amssymb,amsmath]{revtex4}

\begin{document}

\title{Prepotential approach to solvable rational extensions
of Harmonic Oscillator and Morse potentials}

\author{C.-L. Ho}
\affiliation{Department of Physics, Tamkang University,
 Tamsui 251, Taiwan, R.O.C.}

\date{Oct 14, 2011} % {Jun 15, 2011}

\begin{abstract}

We show how the recently discovered solvable rational extensions
of Harmonic Oscillator and Morse potentials can be constructed in
a direct and systematic way, without the need of supersymmetry,
shape invariance, Darboux-Crum and Darboux-B\"acklund
transformations.

\end{abstract}



%%%%%%%%%%%%%%%%%%%%%%%%%%%%%%%%%%%%%%%%%%%
% PACS 2008
% 02.00.00 Mathematical methods in physics
%   02.30.Gp Special functions
%   02.30.Ik Integrable systems
% 03.00.00 Quantum mechanics, field theories, and special relativity
%   03.65.-w Quantum mechanics
%   03.65.Ca Formalism
%   03.65.Db Functional analytical methods
%   03.65.Fd Algebraic methods
%   03.65.Ge Solutions of wave equations: bound states
% 45.00.00 Classical mechanics of discrete systems
%   45.20.-d Formalisms in classical mechanics
%%%%%%%%%%%%%%%%%%%%%%%%%%%%%%%%%%%%%%%%%%%

 \maketitle

\section{Introduction}

It is fair to say that in the last three years some of the most
interesting developments in mathematical physics have been the
discoveries of new types of orthogonal polynomials, called the
exceptional orthogonal polynomials, and the quantal systems
related to them [1-18]. Unlike the classical orthogonal
polynomials, these new polynomials have the remarkable properties
that they still form complete sets with respect to some
positive-definite measure, although they start with degree
$\ell\geq 1$ polynomials instead of a constant.

Two families of such polynomials, namely, the Laguerre- and
Jacobi-type $X_1$ polynomials, corresponding to $\ell=1$, were
first proposed by G\'omez-Ullate et al. in \cite{GKM1}, within the
Sturm-Lioville theory, as solutions of second-order eigenvalue
equations with rational coefficients. The results in \cite{GKM1}
were reformulated in the framework of quantum mechanics in
\cite{Que1}, and in supersymmetric quantum mechanics using
superpotential in \cite{Que2}. These quantal systems turn out to
be rationally extended systems of the traditional ones which are
related to the classical orthogonal polynomials. The most general
$X_\ell$ exceptional polynomials, valid for all integral
$\ell=1,2,\ldots$, were discovered by Odake and Sasaki \cite{OS1}
(the case of $\ell=2$ was also discussed in \cite{Que2}). Later,
in \cite{HOS} equivalent but much simpler looking forms of the
Laguerre- and Jacobi-type $X_{\ell}$ polynomials were presented.
Such forms facilitate an in-depth study of some important
properties of the $X_\ell$ polynomials. Very recently, such
systems have been generalized to multi-indexed cases
\cite{GKM3,OS3}.

Other rational extensions of solvable systems, which are not
related to the exceptional polynomials, are possible
\cite{Har,CPRS,FS,GB,Gran2}. In these systems, the polynomial part
of the wave functions start with degree zero.  One of the simplest
example of such systems was discussed in \cite{CPRS}, which was
later shown to be a certain supersymmetric partner of the harmonic
oscillator in \cite{FS}.  Extending the superpotential scheme of
\cite{FS}, Gandati and B\'erard were able to generate an infinite
set of solvable rational extension for translationally
shape-invariant potentials of the second category \cite{GB}. More
recently, rational extensions of the Morse and Kepler-Coulomb
potentials have also been obtained by means of the
Darboux-B\"acklund transformation in \cite{Gran2}.

From the viewpoint of the generalized Crum's theorem, solvable
rationally extended systems related to the exceptional polynomials
are obtainable from the corresponding ordinary systems, which are
related to the classical orthogonal polynomials, by deleting the
lowest energy levels including the ground states \cite{OS3}.  The
rational extensions of the harmonic and isotonic oscillators
considered in \cite{CPRS,FS,GB} can be obtained in the same way,
with the exception that the ground states were not deleted (see
Appendix A of \cite{GOS}).

So far most of the methods employed to generate solvable rational
extensions of ordinary systems with or without the exceptional
polynomials have invoked in one way or another the ideas of shape
invariance and/or the related Darboux-Crum transformation
(supersymmetry). This requires an exactly solvable ordinary
system, such as the harmonic oscillator, to be known in the first
place, and the superpotential associated with such system is
modified for the extension.

In \cite{Ho2} we have proposed a simple constructive procedure to
generate the exceptional orthogonal polynomials without the need
of shape invariance and Darboux-Crum transformation.  Thus in our
work an exactly solvable ordinary system and its associated
superpotential need not be assumed a priori as in the other works.
The superpotential, as well as the potential, the eigenfunctions
and eigenvalues of the new system are all derived from first
principle in our method. To distinguish the different roles the
superpotential play in our approach and in those employing
Darboux-Crum transformation or supersymmetry, we prefer to call
the superpotential ``prepotential", and our procedure the
``prepotential approach".

It is the purpose of this paper to demonstrate that the solvable
rational extensions of the harmonic oscillator given in
\cite{CPRS,FS,GB} and the Morse potential in \cite{Gran2} can also
be generated very simply in the prepotential approach, without the
need of supersymmetry, shape invariance, Darboux-Crum and
Darboux-B\"acklund transformations.  These two systems are in the
same class as the rationally extended Jacobi system discussed in
Sect. 4.4 of \cite{Ho2}.

\section{Prepotential approach}

The main ideas of the prepotential approach are summarized here.
We refer the reader to \cite{Ho2} for the details of the
procedure. We adopt the unit system in which $\hbar$ and the mass
$m$ of the particle are such that $\hbar=2m=1$.

Consider a wave function $\phi(x)$ which is written in terms of a
function $W(x)$ as $\phi (x)\equiv \exp(W(x))$.  The function
$W(x)$ is assumed to have the form
\begin{eqnarray}
W(x,\eta) = W_0(x)-\ln \xi(\eta) + \ln p(\eta). \label{W}
\end{eqnarray}
Here $\eta(x)$ is a function of $x$ which we shall choose to be
one of the sinusoidal coordinates, i.e., coordinates such that
$\dot{\eta}(x)^2$, where the dot denotes derivative with respect
to $x$, is at most quadratic in $\eta$.

The functions $W_0(x),~\xi(\eta)$ and $ p(\eta)$ are functions to
be determined later. We shall assume $\xi(\eta)$ to be a
polynomial in $\eta$.  The wave function is
\begin{eqnarray}
\phi(x) =\frac{e^{W_0(x)}}{\xi(\eta)}\,p(\eta). \label{phi}
\end{eqnarray}
Operating on $\phi(x)$ by the operator $-d^2/dx^2$ results in a
Schr\"odinger equation ${\mathcal{H}}\phi=0$, where ${\cal{H}}
=-d^2/dx^2 + \bar{V},~~ \bar{V}\equiv  \dot{W}^{2} + \ddot{W}$.
For simplicity of presentation, we shall often leave out the
independent variable of a function if no confusion arises.

Since $W(x)$ determines the potential $\bar{V}$, it is therefore
called the prepotential.  To make $\bar{V}$ exactly solvable, we
demand that: (1) $W_0$ is a regular function of $x$, (2) the
function $\xi(\eta)$ has no zeros in the the ordinary (or
physical) domain of $\eta(x)$, and (3) the function $p(\eta)$ does
not appear in $\bar{V}$.

For $\xi(\eta)=1$, the prepotential approach can generate exactly
and quasi-exactly solvable systems associated with the classical
orthogonal polynomials \cite{Ho3}.  The presence of $\xi$ in the
denominators of $\phi(x)$ and $V(x)$ thus gives a rational
extension, or deformation, of the traditional system.  We
therefore call $\xi(\eta)$ the deforming function.

If the factor $\exp(W_0(x))/\xi(\eta)$ in Eq.~(\ref{phi}) is
normalizable, then $p(\eta)={\rm~ constant}$ (in this case we
shall take $p(\eta)=1$ for simplicity) is admissible.  This gives
the ground state
\begin{eqnarray}
\phi_0(x)=\frac{e^{W_0(x)}}{\xi(\eta)}. \label{phi0}
\end{eqnarray}
However, if $\exp(W_0(x))/\xi(\eta)$ is non-normalizable, then
$\phi_0(x)$ cannot be the ground state. In this case, the ground
state, like all the excited states, must involve non-trivial
$p(\eta)\neq 1$. Typically it is in such situation that the
exceptional orthogonal polynomials arise \cite{Ho2}.  The cases
considered in \cite{Gran2}, which we shall rederived by means of
the prepotential approach in this paper, are such that
$\exp(W_0(x))/\xi(\eta)$ is normalizable and thus $\phi_0(x)$ is
the ground state.

Following the procedure in \cite{Ho2}, we assume $\xi(\eta)$ to
satisfy the equation
\begin{eqnarray}
c_2(\eta)\xi^{\prime\prime} + c_1(\eta)\xi^{\prime} +
\widetilde{\mathcal{E}}(\eta)\xi=0.\label{xi-eq}
\end{eqnarray}
Here the prime denotes derivative with respective to $\eta$. We
choose $c_2(\eta)=\pm \dot{\eta}^2$, and $c_1$ is determined by
\begin{eqnarray}
c_1(\eta)=\pm\left[\frac12\frac{d}{d\eta}\left(\dot{\eta}^2\right)-
2Q(\eta)\right],\label{c1-2}
\end{eqnarray}
where $Q(\eta)\equiv \dot{W}_0\dot{\eta}$. The function
$\widetilde{\mathcal{E}}(\eta)$ was taken to be a real constant in
\cite{Ho2}, but here we allow the possibility that
$\widetilde{\mathcal{E}}(\eta)$ may be a function of $\eta$.
Nonetheless, it turns out that all formulae presented in
\cite{Ho2} remain intact.

By matching Eq.~(\ref{xi-eq}) with the (confluent) hypergeometric
equation, one determines $\tilde{\mathcal{E}},~Q(\eta)$ and
$\xi(\eta)$. Integrating $Q(x)=\dot{W}_0\dot{\eta}$ then gives the
prepotential $W_0(x)$:
\begin{eqnarray}
W_0(x) &=& \int^x dx\, \frac{Q(\eta(x))}{\dot{\eta}(x)}\nonumber\\
&=& \int^{\eta(x)} d\eta\, \frac{Q(\eta)}{\dot{\eta}^2(\eta)};
\label{W0-fm-Q}
\end{eqnarray}
The function $p(\eta)$ is then given by a linear combination of
$\xi$ and $\xi^\prime$:
\begin{eqnarray}
p(\eta)=\xi^\prime(\eta) F(\eta) + \xi(\eta)G(\eta).
\end{eqnarray}
Here the two functions $F(\eta)$ and $G(\eta)$ are determined by
\begin{eqnarray}
F(\eta) &=& c_2(\eta) \cal{V}(\eta), \label{F-V}\\
G(\eta) &=& \left(c_1-c_2^\prime\right){\cal{V}}-c_2{\cal
V}^\prime, \label{G-1}
\end{eqnarray}
with the function $\mathcal{V}(\eta)$ satisfying
\begin{eqnarray}
c_2\mathcal{V}^{\prime\prime} +
\left(2c_2^\prime-c_1\right)\mathcal{V}^\prime
+\left[c_2^{\prime\prime}-c_1^\prime+\widetilde{\mathcal{E}}\pm\mathcal{E}\right]\mathcal{V}=0.
 \label{cal-V}
\end{eqnarray}
By matching Eq.~(\ref{cal-V}) with the (confluent) hypergeometric
equation, one determines $\mathcal{V}$, and thus
$F(\eta),~G(\eta),~p(\eta)$ and $\mathcal{E}$.

Once all the relevant functions and parameters are determined, we
would have constructed an exactly solvable quantal system
$H\phi=\cal{E}\phi$ defined by $H =-d^2/dx^2 + V(x)$, with the
wave function (\ref{phi}) and the potential
\begin{eqnarray}
V(x)&\equiv& \dot{W}_0^{2}+\ddot{W}_0
+\frac{\xi^{\prime}}{\xi}\left[2\dot{\eta}^2\left(\frac{\xi^{\prime
}}{\xi}\right)-\left(2\dot{W}_0\dot{\eta} +\ddot{\eta}\right)\pm
c_1\right]\pm \tilde{\mathcal{E}}.
 \label{Exact-V}
 \end{eqnarray}

Lastly, we note that the functions $p_{\mathcal{E}}$ (here we add
a subscript to distinguish $p$ corresponding to a particular
eigenvalue $\mathcal{E}$) are orthogonal , i.e.,
\begin{equation}
\int d\eta~
p_{\mathcal{E}}(\eta)p_{\mathcal{E}'}(\eta)\frac{\mathcal{W}^2(x(\eta))}{\dot{\eta}}\propto
\delta_{\mathcal{E},\mathcal{E'}} \label{ortho}
\end{equation}
in the $\eta$-space with the weight function
\begin{eqnarray}
\mathcal{W}(x) &\equiv &
\exp\left(\int^x dx\left(\dot{W}_0-\frac{\dot{\xi}}{\xi}\right)\right)\nonumber\\
&=& \frac{e^{W_0(x)}}{\xi_\ell(\eta(x))}. \label{cal-W}
\end{eqnarray}

\section{Harmonic oscillator}

Let us choose $\eta(x)=x \in (-\infty,\infty)$. Then
$\dot{\eta}^2=1$. For $c_2$ and $c_1$, we take the upper signs in
$c_1$ and $c_2$ (it turns out that the lower signs give the same
model) . Thus $c_2(\eta)=1$ and $c_1=-2Q(\eta)$.

Eq.~(\ref{xi-eq}) becomes
\begin{eqnarray}
\xi^{\prime\prime} -2Q(\eta)\xi^\prime + \tilde{\mathcal{E}}
\xi=0. \label{eq-xi-H}
\end{eqnarray}
Comparing Eq.~(\ref{eq-xi-H}) with the Hermite equation
\begin{eqnarray}
H_\ell^{\prime\prime}(\eta) -2\eta H_\ell^{\prime}(\eta) +2\ell
H_\ell(\eta)=0,~~\ell=0,1,2,\ldots, \label{Hermite}
\end{eqnarray}
where $H_\ell(\eta)$ is the Hermite polynomial, we would have
\begin{eqnarray}
\xi(\eta)\equiv
\xi_\ell(\eta;\alpha)=H_\ell(\eta),~~\tilde{\mathcal{E}}=2\ell,
~~Q(\eta)=\eta.
\end{eqnarray}
But this choice is not viable, as
$\xi_\ell(\eta;\alpha)=H_\ell(\eta)$ has zeros in the ordinary
domain $(-\infty,\infty)$, which we want to avoid.  A simple way
to solve this is to make the zeros of Hermite polynomials lie on
the imaginary axis. This is achieved if we set $\eta\to i\eta$ in
Eq.~(\ref{Hermite}), giving
\begin{eqnarray}
H_\ell^{\prime\prime}(i\eta) +2\eta H_\ell^{\prime}(i\eta) -2\ell
H_\ell(i\eta)=0,~~\ell=0,1,2,\ldots. \label{Hermite-I}
\end{eqnarray}
Matching Eq.~(\ref{eq-xi-H}) with (\ref{Hermite-I}) gives
\begin{eqnarray}
\xi(\eta)\equiv
\xi_\ell(\eta;\alpha)=H_\ell(i\eta),~~\tilde{\mathcal{E}}=-2\ell,
~~Q(\eta)=-\eta. \label{xi-H}
\end{eqnarray}
One notes that the Hermite polynomials are odd functions in $\eta$
for odd $\ell$.  So in this case $\xi_\ell$ has a zero at
$\eta=0$.  Further study of this case reveals that the wave
functions are not normalizable. So here we shall only consider the
case with even $\ell=2m (m=1,2,\ldots)$.  By Eq.~(\ref{W0-fm-Q}),
the form of $Q(\eta)$ leads to
\begin{eqnarray}
W_0(x)=-\frac{x^2}{2}. \label{W0-H}
\end{eqnarray}
We shall ignore the constant of integration as it can be absorbed
into the normalization constant. As noted in Sect.~II, in this
case $p(\eta)=1$ is admissible, as $\exp(W_0(x))/\xi(\eta)$ is
normalizable.  So the energy and eigenfunction of the ground state
of this system are ${\cal E}_0=0$ and
$\phi_0(x)=\exp(W_0(x))/\xi(\eta)$.  Below we determine the
energies and eigenfunctions of the excited states.

With the solutions in Eq.~(\ref{xi-H}), Eq.~(\ref{cal-V}) becomes
\begin{eqnarray}
\mathcal{V}^{\prime\prime} -2\eta\mathcal{V}^\prime
+\left(\mathcal{E}-2\ell-2\right)\mathcal{V}=0.
 \label{cal-V-H}
\end{eqnarray}
Comparing Eqs.~(\ref{cal-V-H}) and (\ref{Hermite}) (with $\ell$
replaced by $n=0,1,2,\ldots$), we get
\begin{eqnarray}
\mathcal{V}(\eta)=H_n(\eta),~~\mathcal{E}\equiv
\mathcal{E}_{\ell,n}=2(n+\ell+1),~~\ell=2m. \label{H-V-E}
\end{eqnarray}
From Eqs.~(\ref{F-V}) and (\ref{G-1}), one eventually obtains
\begin{eqnarray}
p(\eta)&\equiv & p_{\ell,n}(\eta)=\xi^\prime F+ \xi G\nonumber\\
&=&H_n(\eta)\xi_\ell^\prime(\eta) + \left[2\eta H_n(\eta)-
H_n^\prime(\eta)\right]\xi_\ell(\eta)\nonumber\\
&=& H_n(\eta)H_\ell^\prime(i\eta) + H_{n+1}(\eta)H_\ell(i\eta).
\label{P-H}
\end{eqnarray}
Use has been made of the identity $H_n^\prime=2\eta H_n-H_{n+1}$
in obtaining the last line in Eq.~(\ref{P-H}). We note that
$p_{\ell,n}(\eta)$ is a polynomial of degree $\ell+n+1$.

By Eq.~(\ref{ortho}), one finds that $p_{\ell,n}(\eta;\alpha)$'s
are orthogonal in the sense
\begin{equation}
\int^\infty_{-\infty}\,d\eta~\frac{e^{-\eta^2}}{\xi_\ell^2}\,
p_{\ell,n}(\eta;\alpha)p_{\ell,k}(\eta;\alpha)\propto \delta_{nk}.
\end{equation}

The exactly solvable potential is given by Eq.~(\ref{Exact-V})
with $W_0(x)$ and $\xi_\ell(\eta;\alpha)$ given by
Eqs.~(\ref{W0-H}) and (\ref{xi-H}), respectively. Explicitly, the
potential is
\begin{equation}
V(x)=x^2 -1
+2\frac{\xi_\ell^{\prime}}{\xi_\ell}\left[\left(\frac{\xi_\ell^{\prime}}
{\xi_\ell}\right)+ 2\eta\right]+ 2\ell.
\end{equation}
The complete eigenfunctions and energies are
\begin{eqnarray}
\phi_0 (x;\alpha) &\propto &
\frac{e^{-\frac{x^2}{2}}}{\xi_\ell},~~~~~~~~~~~~~~~~~~\mathcal{E}_0=0,\\
\phi_{\ell,n}(x;\alpha) &\propto &
\frac{e^{-\frac{x^2}{2}}}{\xi_\ell}p_{\ell,n}(\eta(x);\alpha),~~\mathcal{E}_{\ell,n}=2(n+\ell+1),~~n=0,1,2\ldots.
\label{phi-H}
\end{eqnarray}
Using the identity for $\ell=2m$ \cite{Szego,Magnus}, i.e.,
\begin{eqnarray}
H_{2m}(i\eta)=(-1)^m 2^{2m}m! L_m^{(-\frac12)}(-\eta^2),
\end{eqnarray}
where $L_\ell^{(\alpha)}(\eta)$ is the Laguerre polynomial, and
the identity
\begin{equation}
\frac{d}{d\eta}L_\ell^{(\alpha)}(\eta)=-L_{\ell-1}^{(\alpha+1)}(\eta),
\label{L1}
\end{equation}
we can reduce Eq.~(\ref{phi-H}) to
\begin{eqnarray}
\phi_{2m,n}\sim
\frac{e^{-\frac{x^2}{2}}}{L_m^{(-\frac12)}(-\eta^2)} \left[\frac12
L_m^{(-\frac12)}(-\eta^2) H_{n+1}(\eta) + \eta
L_{m-1}^{(\frac12)}(-\eta^2) H_n(\eta)\right].
\end{eqnarray}
This result is identical with that given in \cite{Gran2}.

\section{Morse potential}

Now we consider rational extension of the Morse potential.  It
turns out that in this case $\tilde{\mathcal{E}}$ cannot be a
constant.

Let us choose $\eta(x)=e^{-x}\in (0,\infty)$, with
$\dot{\eta}^2=\eta^2$. For definiteness we shall take the upper
signs for $c_2$ and $c_1$, as the lower signs lead to the same
results.  So we have $c_2(\eta)= \eta^2$ and
$c_1=(\eta-2Q(\eta))$.

Equation determining $\xi$ is
\begin{eqnarray}\eta^2\xi^{\prime\prime}(\eta)
+\left(\eta-2Q(\eta)\right)\xi^\prime (\eta)+
{\tilde{\mathcal{E}}}(\eta) \xi(\eta)=0.\label{eq-xi-M}
\end{eqnarray}
In order to link Eq.~(\ref{eq-xi-M}) with the Laguerre equation
\begin{eqnarray}
\eta L_\ell^{\prime\prime(\alpha)} + \left(\alpha +1 -\eta
\right)L_\ell^{\prime(\alpha)} +\ell
L_\ell^{(\alpha)}=0,~~\ell=0,1,2,\ldots, \label{Lag}
\end{eqnarray}
we rewrite Eq.~(\ref{eq-xi-M}) as
\begin{eqnarray}
\eta\xi^{\prime\prime}(\eta) +
\left(1-2\frac{Q(\eta)}{\eta}\right)\xi^\prime
(\eta)+\frac{{\tilde{\mathcal{E}}(\eta)}}{\eta}
\xi(\eta)=0.\label{eq-xi-M-1}
\end{eqnarray}
Directly matching this equation with Eq.~(\ref{Lag}) will leads to
unnormalizable wave functions.  So instead we set $\eta\to -\eta$
in Eq.~(\ref{eq-xi-M-1}).  This gives
\begin{eqnarray}
\eta\xi^{\prime\prime}(-\eta) +
\left(1+2\frac{Q(-\eta)}{\eta}\right)\xi^\prime
(-\eta)+\frac{{\tilde{\mathcal{E}}(-\eta)}}{\eta}
\xi(-\eta)=0.\label{eq-xi-M-2}
\end{eqnarray}
Comparing this equation with Eq.~(\ref{Lag}) leads to
\begin{eqnarray}
\xi(-\eta)\equiv
\xi_\ell(-\eta;\alpha)=L_\ell^{(\alpha)}(\eta),~~\tilde{\mathcal{E}}(-\eta)=\ell\eta,~~
Q(-\eta)=\frac{\alpha}{2}\eta-\frac12\eta,
\end{eqnarray}
or equivalently,
\begin{eqnarray}
\xi_\ell(\eta;\alpha)=L_\ell^{(\alpha)}(-\eta),~~\tilde{\mathcal{E}}(\eta)=-\ell\eta,~~
Q(\eta)=-\frac{\alpha}{2}\eta-\frac12\eta, \label{xi-M}
\end{eqnarray}
The form of $Q(\eta)$ leads to
\begin{equation}
W_0(x)=-\frac{\alpha}{2}\ln\eta - \frac{\eta}{2}.
\end{equation}
According to the Kienast-Lawton-Hahn's Theorem
\cite{Szego,Magnus}, the deforming function $\xi_\ell(\eta)$ will
have no positive zeros in $(0,\infty)$ if: (i) $-2k-1<\alpha <-2k$
with $-\ell<\alpha <-1$, or (ii)  $\ell$ is even with $\alpha
<-\ell$.  Again, in this case, $p(\eta)=1$ is admissible.  Thus
the energy and eigenfunction of the ground state of this system
are ${\cal E}_0=0$ and $\phi_0(x)=\exp(W_0(x))/\xi(\eta)$.  We now
determine the energies and eigenfunctions of the excited states.

With the solutions in Eq.~(\ref{xi-M}), Eq.~(\ref{cal-V}) becomes
\begin{eqnarray}
\mathcal{V}^{\prime\prime}
+\left(-\alpha+3-\eta\right)\mathcal{V}^\prime
+\left[\frac{\mathcal{E}-\alpha+1}{\eta}-(\ell+2)\right]\mathcal{V}=0.
 \label{cal-V-M}
\end{eqnarray}
In order that $\mathcal{E}$ be dependent on $n$, we try
$\mathcal{V}=\eta^\gamma U(\eta)$ where $\gamma$ is a real
parameter and $U(\eta)$ a function of $\eta$.  From
Eq.~(\ref{cal-V}) we get
\begin{eqnarray}
\eta U^{\prime\prime} + \left(2\gamma-\alpha +3 -\eta
\right)U^{\prime}
+\left[\frac{\mathcal{E}-\alpha+1+\gamma(\gamma-\alpha+2)}{\eta}-(\gamma+\ell+2)\right]U=0.
\label{U-M}
\end{eqnarray}
Matching this equation with Eq.~(\ref{Lag}), we have
($n=0,1,2,\ldots$)
\begin{eqnarray}
\gamma &=& -(n+\ell+2),\nonumber\\
\mathcal{E}_{\ell,n}&=&
\alpha-1-\gamma(\gamma-\alpha+2)=\alpha-1-(n+\ell+2)(n+\ell+\alpha),\label{E-Morse}\\
\beta&=&2\gamma-\alpha+2=-\alpha-2(n+\ell+1),\nonumber\\
 U_n(\eta)&=&L_n^\beta(\eta),
~~\beta>-1.
\nonumber
\end{eqnarray}

Putting all these results into $F(\eta)$ and $G(\eta)$ gives
\begin{eqnarray}
p(\eta)&\equiv&
p_{\ell,n}(\eta;\alpha)=\eta^{-n-\ell-1}P_{\ell,n}(\eta;\alpha)\nonumber\\
P_{\ell,n}(\eta;\alpha)&\equiv & \eta L_n^{(\beta)}\xi_\ell^\prime
-\left(\ell L_n^{(\beta)}+ (n+1) L_{n+1}^{(\beta)}\right)\xi_\ell.
\label{P-M}
\end{eqnarray}
$P_{\ell,n}(\eta;\alpha)$ is a polynomial of degree $\ell+n+1$. It
is also easy to check that $p_{\ell,n}(\eta;\alpha)$'s are
orthogonal with respect to the weight function
\begin{equation}
\frac{e^{-\eta}\eta^{-\alpha}}{\xi_\ell^2}.
\end{equation}

The exactly solvable potential is given by
\begin{equation}
V(x)=\frac{1}{4}e^{-2x}+
\frac12(\alpha-4\ell-1)e^{-x}+\frac{\alpha^2}{4}
+2\frac{\xi_\ell^{\prime}}{\xi_\ell}\left[e^{-2x}\left(\frac{\xi_\ell^{\prime}}{\xi_\ell}+1\right)+
\alpha e^{-x}\right].
\end{equation}
The complete eigenfunctions are
\begin{eqnarray}
\phi_0(x,\alpha) &\propto &
\frac{e^{-\frac{\eta}{2}}\eta^{-\frac{\alpha}{2}}}{\xi_\ell},\\
\phi_{\ell,n}(x;\alpha) &\propto &
\frac{e^{-\frac{\eta}{2}}\eta^{-\frac{\alpha}{2}}}{\xi_\ell}p_{\ell,n}(\eta;\alpha),
\end{eqnarray}
where $p_{\ell,n}(\eta;\alpha)$'s are given in (\ref{P-M}).  The
corresponding eigen-energies are $\mathcal{E}_0=0$ and
$\mathcal{E}_{\ell,n}$ in (\ref{E-Morse}), respectively. For the
wave functions to be regular at $x=0$, one must have $n<-\alpha/2
-\ell-1$.  This means the system admits only finite number of
bound states.

With the help of the identities (\ref{L1}) and
\begin{eqnarray}
L_\ell^{(\alpha)}(\eta)-L_\ell^{(\alpha-1)}(\eta)&=&L_{\ell-1}^{(\alpha)}(\eta),
\label{L-2}\\
\eta L_{\ell-1}^{(\alpha+1)}(\eta) -\alpha
L_{\ell-1}^{(\alpha)}(\eta)&=&-\ell L_\ell^{(\alpha-1)}(\eta),
\label{L-3}
\end{eqnarray}
one can recast $P_{\ell,n}(\eta;\alpha)$ in (\ref{P-M}) into
\begin{eqnarray}
P_{\ell,n}(\eta;\alpha)= -\left[\left(\alpha+\ell\right)
L_{\ell-1}^{(\alpha)}(-\eta)L_n^{(\beta)}(\eta) + (n+1)
L_{\ell}^{(\alpha)}(-\eta)L_{n+1}^{(\beta)}(\eta)\right].
\label{P-M-1}
\end{eqnarray}
This expression is exactly the same as that given in \cite{Gran2}
in the case $\ell=2m$, with the identification
\begin{eqnarray}
\alpha &=&-2(a+\ell+1),\nonumber\\
n &\to& k,\\
\beta &=& -\alpha-2(n+\ell+1)=2(a-k).
\end{eqnarray}

\section{summary}

We have shown how the recently discovered solvable rational
extensions of Harmonic Oscillator and Morse potentials can be
constructed in a direct and systematic way, without the need of
supersymmetry, shape invariance, Darboux-Crum and
Darboux-B\"acklund transformations.  In our approach,  the
prepotential, the deforming function, the potential, the
eigenfunctions and eigenvalues are all derived within the same
framework.

With the results given here and in \cite{Ho2}, rational extensions
of all well-known one-dimensional solvable quantal systems based
on sinusoidal coordinates have been generated  by the prepotential
approach.  One would like to apply the same approach to find
rational extensions of the other solvable models based on
non-sinusoidal coordinates, following the work of the third paper
in \cite{Ho3}. Unfortunately, such way of rational extensions only
lead to quasi-exactly solvable systems, because the energy quantum
number $n$ will appear in the $x$-dependent terms in $V(x)$.  A
non-trivial generalization of the present approach may be in
order, which we hope to report in the near future.

\section*{Acknowledgments}

This work is supported in part by the National Science Council
(NSC) of the Republic of China under Grant NSC
NSC-99-2112-M-032-002-MY3.

\end{document}